# Criterion for sliding / rolling characterization during droplet motion over superhydrophobic surfaces

*Indrajit P. Wadgaonkar[1a], T.Sundararajan[1], Sarit K.Das[1b]* *

[1]*Department of Mechanical Engineering, IIT Madras, Chennai-600036, India.*

## ABSTRACT

Superhydrophobic surfaces have been the focus of research in the recent years. One of the reasons for this is the self cleaning property of these surfaces which emerges from the ability of the droplets to roll freely over them. However majority of the studies available in literature are on the static wetting behavior of liquid droplets on such surfaces and the physics of the motion of droplets has not been studied exhaustively either theoretically or experimentally. In the present study droplet motion on superhydrophobic surfaces has been modeled to analyze the sliding/ rolling characteristics of the droplet motion. A non-dimensional number is proposed to indicate whether a given droplet would tend to roll or slide more on a given superhydrophobic surface. We refer to this number as 'Slip Reynolds' number. Simulations of droplet motion were carried out with different surface and droplet characteristics leading to a unique value of this number

[a] This research was performed when Indrajit P. Wadgaonkar was at IIT Madras, Chennai-600036, India.

[b] Author to whom correspondence should be addressed. Electronic mail: skdas@iitm.ac.in (S.K. Das)

corresponding to sliding/ rolling behavior. The applicability of this number was tested with available experiments from the literature.

## 1. INTRODUCTION:

Superhydrophobic surface flows have been a prime area of fluid dynamics research in the recent years. Their potential in varied applications stems from the fact that droplets roll off the surfaces due to low wettability. However, the physics of the motion of droplets over superhydrophobic surfaces has not been understood comprehensively yet. And hence the sliding/rolling characteristics of the droplet motion on these surfaces, which are extremely important with respect to the applications, remain largely unaddressed. On the theoretical side, the appearance of varied length scales during the movement of contact line has been one of the prime reasons for the lack of a convincing analytical model so far. Phenomena such as dynamic contact angle variation with the velocity of the contact line and occurrence of unequal contact angles at the advancing and the receding ends further complicate the physics. Earlier researchers largely focused on the theoretical study of movement of the contact line on a plane surface that gives rise to dynamic contact angle variation *(1-5)*. Recently, Shikhmurzaev *(6, 7)* has proposed a theory that tries to explain the contact line motion from first principles. Although Monnier and Witomski *(8)* and Billingham *(9)* have tried to implement Shikhmurzaev model for a receding film, there is not widespread implementation of this model as yet. An empirical relation between the dynamic contact angle and the contact line velocity has been derived experimentally by Hoffman *(10)*.

With regard to the study of droplet motion over superhydrophobic surfaces, another important aspect is the transition from Cassie to Wenzel state *(11-15)*. All the work in this area is focused on a static drop over superhydrophobic surface. The expressions for the free energy of a static drop, the minimization of which decides its equilibrium state (Patankar *(12)*, Marmur *(13)* and others) reveal that whichever state has a lower equilibrium contact angle is the state with the lower Gibb's free energy *(11)* and hence is the preferred state. However, Patankar *(13)* argues that for a superhydrophobic surface, an energy barrier exists between the energies of the Cassie and Wenzel states and this energy barrier is higher than the individual energies of Cassie and Wenzel states and hence work is to be supplied for the transition even if the Wenzel state is the preferred one. Other works are those of Bormashenko et al. *(14)* who studied the vibration induced transition from Wenzel to Cassie state and Lafuma and Querre *(15)* who present a good review on the wetting and roughness characteristics of a surface and arrive at a similar expression for energy barrier for Cassie to Wenzel transition as that of Patankar *(13)*.

Another possible way to reveal the underlying physics would have been simulating the moving interface problem numerically, using the known equations of fluid flow and empirical relations for the dynamic contact angle. However, in problems involving moving interfaces, parasitic currents pose major difficulties in numerical simulations while using the CSF (Continuum Surface Force) model for surface tension given by Brackbill et al. *(16)*. The CSF model does not converge to more accurate results with mesh refinement (Bohacek *(17)*); in fact, parasitic currents tend to increase with mesh refinement on fine meshes and are inversely proportional to the capillary number (Harvie et al. *(18)*). Out of the few numerical studies on the moving contact line problem, are those of Dupont and Legendre *(19)*, Dupius and Yeomans *(20)* and Renardy et al. *(21)* who have simulated the moving contact line problem numerically, based on techniques such as Lattice Boltzmann Method or Volume of Fluid model. The approach Sikalo *(22)* appears apt, where the region very close to the contact line is replaced by a dynamic contact angle boundary condition at the wall, with the dynamic contact angle given as a function of the contact line velocity. The flow in the domain is solved by the traditional Navier-Stokes equations. Although the flow dynamics very close to the contact line are neglected in this approach, nonetheless it could be used to reveal vital physics on the scale of the droplet radius.

Experimentally it is extremely difficult to resolve the motion of the contact line. Most of the available experimental studies have focused on the measurement of the dynamic contact angle (Dorrer and Ruhe *(23)*) or the sliding angle (Yoshimitsu et al. *(24)*, Lv et al. *(25)*). One of the experimental works revealing the sliding / rolling characteristics of drop motion on superhydrophobic surfaces was that of Hao et al. *(26)* who conducted a PIV study of the motion of a 5 µL drop on a super hydrophobic surface made of slender pillars and the internal velocity profiles in a 2D plane sliced for the PIV study were reported**.** It was found that rolling occurred only on the edge of the drop and in the middle portion sliding prevailed. Viscosity of the drop fluid determines the extent to which rolling affects the drop. However, in another experimental work, Richard and Quere *(27)* reported that the drop rolls with negligible sliding. Sakai et al. *(28, 29)* obtained the images of drop shapes on a hydrophobic surface and their plots of average drop velocity v/s time reveal the sliding / rolling characteristics to some extent.

Thus, the available literature shows that the current theoretical and experimental findings are far from giving a complete picture of the phenomenon of droplet motion over super hydrophobic surfaces and hence the underlying physics remains largely unexplored. Moreover, majority of works have been on static drops only, thus revealing very little information about the dynamic state which in fact is required from the application point of view for a superhydrophobic surface. While some experiments suggest sliding behavior of drops *(26, 29)*, others deem it to be rolling *(23, 27)* and there is no theoretical means to determine whether a given drop would roll or slide on a given surface. In the present study, we try to answer this very question about the mode of droplet motion on a superhydrophobic surface. Our aim is to develop a theoretical criterion to predict whether the droplet would slide more or roll more on a given superhydrophobic surface and demonstrate the uniqueness of the criterion. Since the requirement of high computational time has limited the overall scope of numerical simulations, we use the conventionally less expensive VOF method to model droplet motion instead of more expensive techniques like Lattice Boltzmann or molecular dynamics. However, special care has been taken to check the accuracy of the numerical technique and its validity with respect to critical experimental observations. The approach of Sikalo *(22)* has been used in the present study to simulate droplet motion. Since all the empirical relations used to express the dynamic contact angle as a function of contact line velocity have their origin in the experiments conducted for droplet motion on a flat surface, whether or not these remain valid for droplet motion on super hydrophobic surfaces is a question which cannot be answered with certainty. Therefore in the present study, the expression derived on the basis of Shikhmurzaev's *(6, 7)* theory for the dynamic contact angle has been used to impose the dynamic contact angle at the wall. From the way the theory has been formulated, it should work equally well for droplet motion over the superhydrophobic surfaces also. The surface parameters are chosen from the available literature *(11-15)* so that the droplet can be expected to remain in the Cassie state throughout its motion.

## 2. NUMERICAL FORMULATION

### 2.1: Volume of Fluid method

In the present study, the Volume Of Fluid (VOF) method has been used to simulate two phase motion of water and air in 2D. The motion of contact line has been analyzed previously by Cox *(1),* Hocking *(3)*, Dussan and Davis *(2)*, Shikhmurzaev *(6, 7)* and others in 2D. Also, the available PIV studies on drop motion have reported the features of drop motion in a 2D plane.

Moreover, when the ratio between the typical drop size and the typical heterogeneity scale is large, 3D effects are small in the scale of the heterogeneity *(12)*. In the present simulations, the ratio between drop size and heterogeneity scale is more than 10 and hence the results obtained in the present 2D study are expected to adequately and closely represent the actual 3D scenario *(12)*. The bulk flow is governed by the Navier Stokes equations while the interface is tracked by the VOF model. The simulations have been carried out using the commercial CFD software Fluent 6.3.26, including user defined functions (UDF) for the dynamic contact angle variations.

## 2.2. Modeling of Surface Tension Force

The surface tension force is modeled by the continuum surface force (CSF) model proposed by Brackbill et al. *(16)* which expresses the force at the surface as a volume force using the divergence theorem. It is this volume force that is added to the momentum equation as a source term in the form

$$F_{vol} = \sigma_{ij} \frac{\rho \kappa_i \nabla \alpha_i}{\frac{1}{2}(\rho_i + \rho_j)} \tag{1}$$

Where $\sigma_{ij}$ is the surface tension between the two fluids (air and water in the present study), $\rho$ is the volume averaged density in each cell, $\rho_i$ and $\rho_j$ are the densities of the two phases (air and water in the present study), $\kappa_i$ is the local interfacial curvature and $\nabla \alpha_i$ is the gradient of the volume fraction for species 'i'.

## 2.3. Dynamic contact angle

The dynamic contact angle has been supplied as a boundary condition at the wall with the help of User Defined Functions (UDF) in Fluent. The expression derived by Shikhmurzaev *(6)* relating the dynamic contact angle with the velocity of the contact line, has been used as follows:

$$cos\theta_s - cos\theta_d = \frac{p_{SG}^s + cos\theta_s + (1 - \rho_{1e}^s)^{-1}[1 + \rho_{1e}^s u_o(\theta_d)]}{1 + (2V)^{-1}[(V^2 + 1)^{\frac{1}{2}} - V]} \tag{2}$$

Where, $\theta_s$ is the static contact angle and $\theta_d$ is the dynamic contact angle to be determined. Further, following Shikhmurzaev *(6)* $p^s_{SG}$ is defined as the non-dimensionalised surface pressure between gas and solid (taken to be zero here) and, $\rho_{1e}{}^s$ is the equilibrium non-dimensionalised surface density. The other terms in equation (2) are defined as,

$$u_o(\theta_d) = \frac{sin\,\theta_d - \theta_d\,cos\,\theta_d}{sin\,\theta_d\,cos\,\theta_d - \theta_d} \;\; \text{and} \; V = \frac{\tau\beta\,U^2}{\sigma\lambda\,(1+4A)} \qquad (3)$$

In the above equation, V represents the dimensionless speed of contact line movement and U is the dimensional contact line velocity determined from the numerical solver. Based on an order of magnitude analysis, Shikhmurzaev *(7)* has shown that $\lambda$ can be simplified as, $\lambda = \frac{1}{(1-\rho^s_{1e})}$. For the flow situation under consideration the values of the constants are given as $\lambda$=20, $\rho_{1e}{}^s$=0.95 and A=1/12. The relaxation time $\tau$ was taken as $10^{-8}$ s and the co-efficient of sliding friction, $\beta$ was taken as $10^7$ kg/ m$^2$s *(9)*. Also $\sigma$ is the surface tension co-efficient between water and air. It is seen that the above equation for $\theta_d$ is transcendental in nature and hence it is solved by Newton's Bisection method. As mentioned in *(6)*, the equation for dynamic contact angles is only valid for capillary number, Ca << 1. Since $Ca = \mu U / \sigma$, substituting for $\mu$ and $\sigma$ for water at room temperature and typical values of contact line velocities ( U ~ 0.1 m/s), it can be shown that the use of eq.(2) for dynamic contact angle is justified.

## 2.4. Numerical Technique and Grid Independence

Since the numerical scheme uses Brackbill's Continuum Surface Force (CSF) algorithm to model the surface tension force; the simulations are restricted by the capillary time step as given by Brackbill et al. *(16)* as:

$$\Delta t \approx \sqrt{\frac{\rho\,\Delta x^3}{2\pi\sigma}} \qquad (4)$$

The mesh size for the present simulation was 2 microns and hence as per the above equation, $\Delta t \approx 10^{-7}$ s.

Galusinski and Vigneaux *(30)* claimed a new stability condition on time step for flows with low and medium Reynolds numbers. This condition is at least one order more restrictive than the capillary time step given above and has the following form:

$$\Delta t \approx \frac{\mu \Delta x}{\sigma} \qquad\qquad (5)$$

For the present simulations it works out to be: $\Delta t \approx 2 \times 10^{-8}$ s.

Indeed during computations, a time step of $10^{-8}$ s was found to provide good convergence characteristics, in general. However, due to the non-linearity of the problem, the shape of the droplet as it moves over the posts was found to be better captured for a time step of $10^{-9}$ s and hence the time step for the present simulations was maintained at $10^{-9}$ s. Usage of such low time steps poses a serious problem with regard to the overall computational time required and hence simulations take a long time. The flow equations are solved on a double precision solver using the finite volume method. Suitable body forces have been imposed depending on the parameters of the respective simulations such as the area of solid-liquid contact, droplet radius and fluid properties like viscosity, etc. The following logic is implemented to track the front and rear ends of the droplet to specify the advancing and the receding contact angles in the UDF:

When $\alpha$ is the volume fraction of the secondary phase,

$\nabla \alpha < 0$ at the boundary cell , the cell is on the receding side and,

$\nabla \alpha > 0$ at the boundary cell, the cell is on the advancing side.

It was found in the course of the simulations that the spreading characteristics and hence the shape of the droplet vary with the mesh spacing; this is akin to the fact that wettability is affected by the nature of the surface. Hence grid independence was carried out considering the shape of the droplet at a corresponding position, for various grid spacings as well as average velocity values of the droplet. Uniform grids having a spacing of 1, 1.6, 2 and 4 microns were tried out. As seen from Fig.1 the shape deteriorates as the grid is refined too much, contrary to the general belief that finer grids lead to more accurate results. The reason was found to be the dominance of spurious currents with grid refinement *(17)*. As seen from the streamlines shown in Fig.1, with decreasing mesh size, the stream lines exhibit more discontinuity which is a clear indication of the growing dominance of spurious currents.

**Table I.Average velocity of droplet**

| Mesh size (microns) | Average Velocity(m/s) | Pressure variation near corners x $10^2$ (Pa) |
|---|---|---|
| 4 | 1.33 | 9-14 |
| 2 | 0.665 | 8-14 |
| 1.6 | 0.665 | 11-18 |
| 1 | 0.562 | 29-35 |

The sharp rise in the maximum pressure with grid size of 1 micron (as seen from Table I) is also attributable to these spurious currents. Hence, in the present simulations, it was decided to use a suitable grid size which is just enough to capture the interfacial shape properly while not introducing excessive spurious currents at locations of sharp interfacial shape change. The results for grid sizes of 2 micron and 1.6 micron were found to match reasonably well as seen from Fig.1 and Table I. Also the effect of spurious currents was found to be negligible for these meshes throughout the flow domain as seen from the streamlines and hence a uniform grid of 2 micron was used for all the subsequent simulations. The grid size of 4 microns x 4 microns shows less influence of parasitic currents but it is too coarse to capture the interface shape well and hence it is not the optimum.

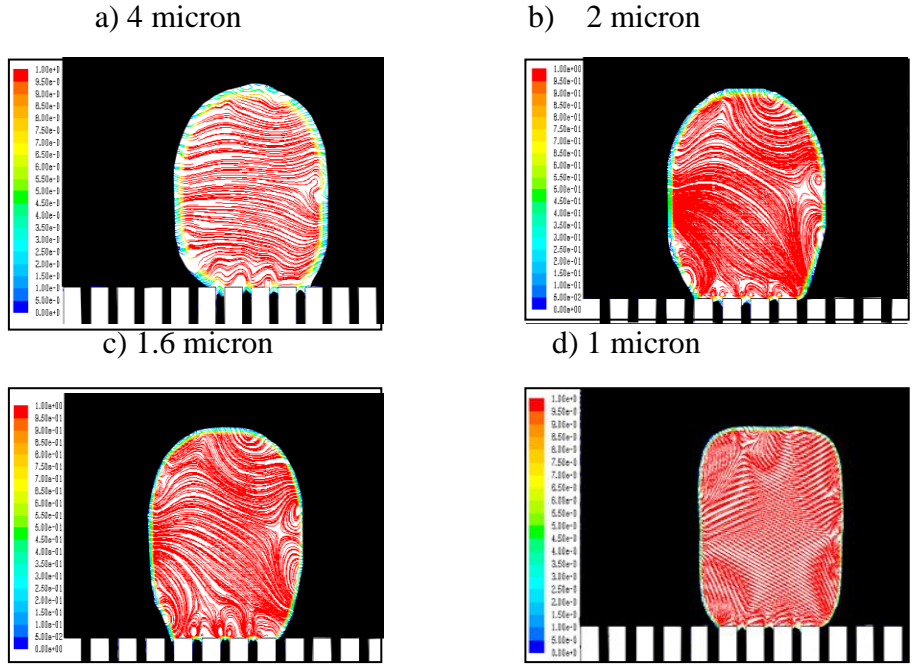

**FIG.1.** Streamlines for various grid sizes

## 3. RESULTS AND DISCUSSION

Simulations were first carried out for drop motion on a flat plate. The shape of the drop at steady state was compared to that predicted by Dupont and Legendre *(19)*. Also simulations were carried out for drop motion on a hydrophobic surface and the drop shape was compared with the experimental work of Sakai et al. *(28)*. Figure 2 shows that the drop shapes from numerical simulations are in close agreement with those of Lattice Boltzmann predictions of Dupont and Legendre *(19)* as well as the experimental work of Sakai et al. *(28)*. Hence the numerical model can be taken as adequately validated.

### 3.1 Droplet motion over superhydrophobic surfaces

Following were the parameters selected for the simulations:

Droplet radius = 166 microns

Post width, d=16 microns; Post height, h=40 microns.

Post spacing, s= 2 microns, 4 microns and 8 microns respectively for successive simulations.

From the available experimental works *(23, 24, and 25)* in literature for static drops, this particular geometry is expected to retain the Cassie state throughout the simulations which is extremely important to study the dynamic characteristics of drop motion. For the computational domain, no slip boundary condition was imposed on the bottom wall composed of posts and outflow boundary condition was imposed on the other surfaces.

a)

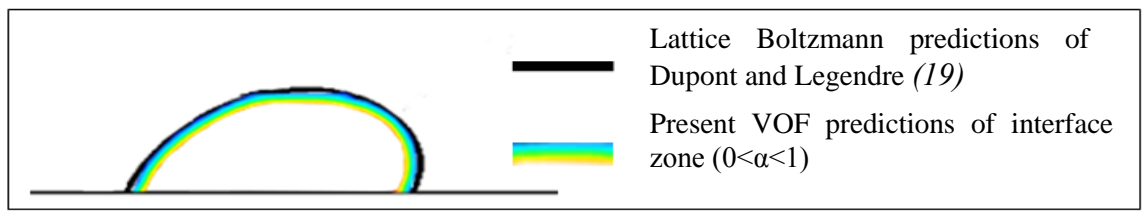

Lattice Boltzmann predictions of Dupont and Legendre *(19)*

Present VOF predictions of interface zone (0<α<1)

b)

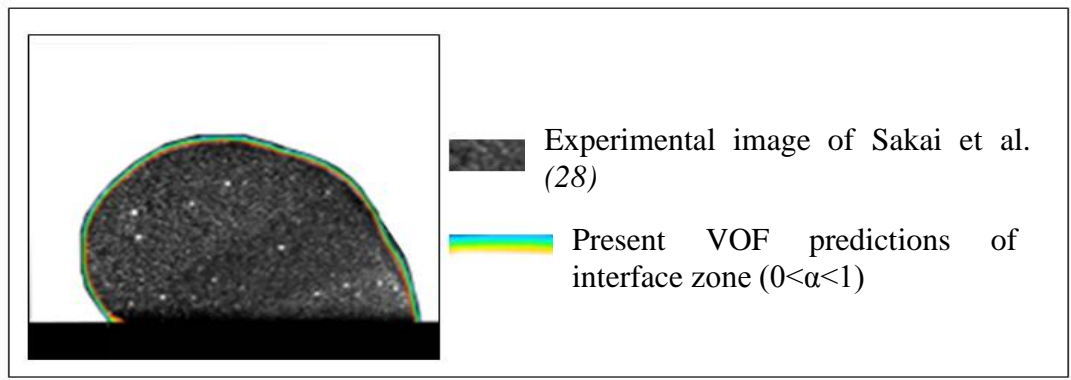

Experimental image of Sakai et al. *(28)*

Present VOF predictions of interface zone (0<α<1)

**FIG.2.** a) Comparison of droplet profiles as predicted by Dupont and Legendre *(19)* and present simulation b) Validation of volume fraction contours with experimental work of Sakai et al. *(28)*.

Figures 3a, 3b and 3c shows the contour plots of volume fraction for the surface geometries with 2 micron, 4 micron and 8 micron post spacing, respectively.

a)

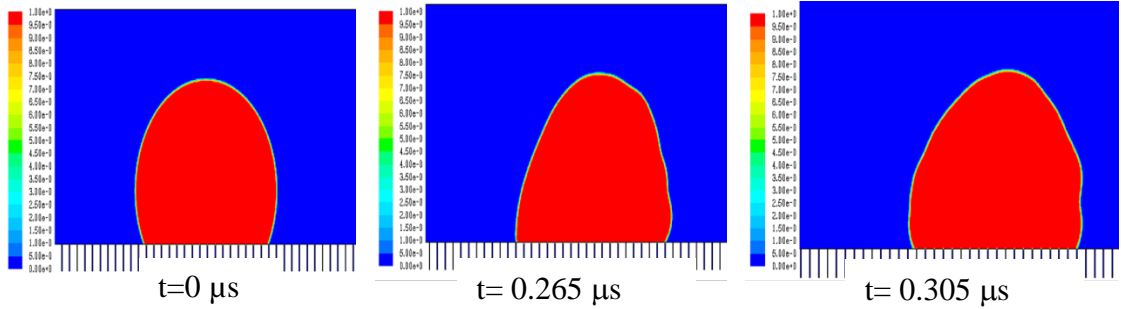

t=0 µs          t= 0.265 µs          t= 0.305 µs

b)

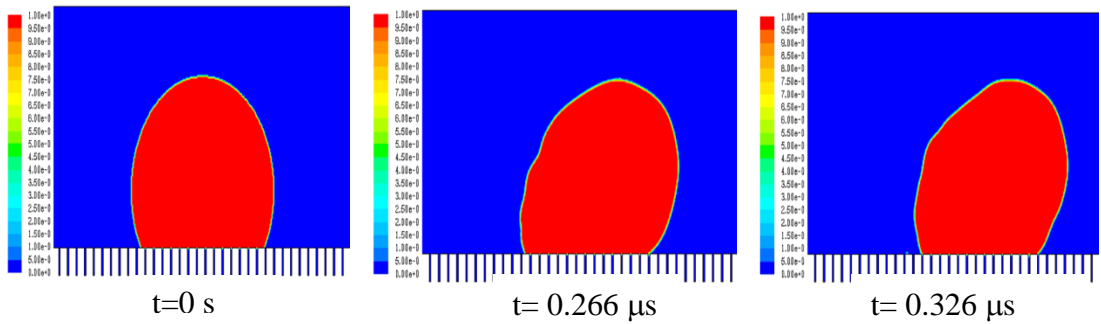

t=0 s          t= 0.266 µs          t= 0.326 µs

c)

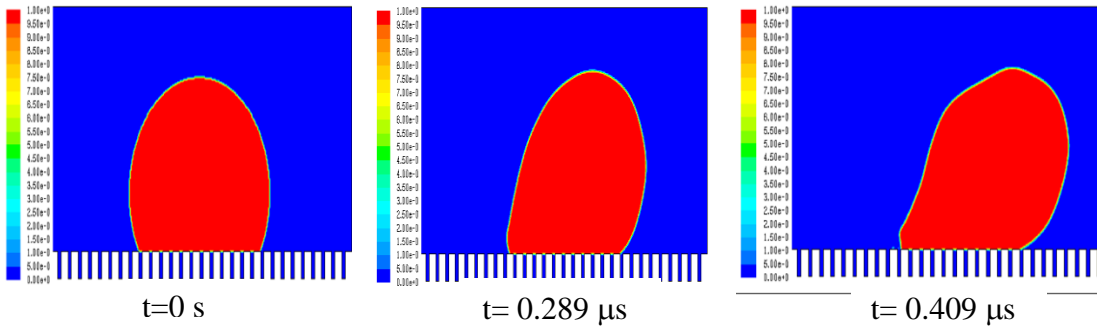

t=0 s          t= 0.289 µs          t= 0.409 µs

**FIG.3.** Droplet motion for Post width, d=16 micron; Post height, h=40 micron; and Post spacing, a) s= 2 microns, b) s=4 microns and c) s=8 microns.

An animation clip has been provided, Fig. 4 (Multimedia view) depicting the drop deformation from its initial state and its subsequent motion over the superhydrophobic surface with the following parameters: Post width, d=16 microns; Post height, h=40 microns; Post spacing, s= 4 microns; Droplet radius = 166 microns. Stick slip phenomenon was observed for the droplet as it moves over the posts. This mechanism of droplet movement on superhydrophobic surfaces is consistent with that predicted by Dorrer and Ruhe *(23)* and Lv et al. *(25)*. The advancing end movement was analyzed in detail. The advancing end gets pinned with the corresponding contact angle increasing gradually till the advancing front touches the subsequent post. After that the droplet starts moving on the post surface. This was consistent with the results from the Lattice Boltzmann simulations of Zhang et al. *(31)*. (Details of the motion of advancing front of the droplet are given in the appendix). From the observed motion of the droplet over superhydrophobic surfaces it was found that the contact line velocity is very small and hence from eq.(2), $\theta_d \approx \theta_s$. $\theta_s$ is the static equilibrium contact angle taken from literature *(26)*. Thus although dynamic contact angle boundary condition was imposed in the simulations, for droplet motion on superhydrophobic surfaces it is of very little significance.

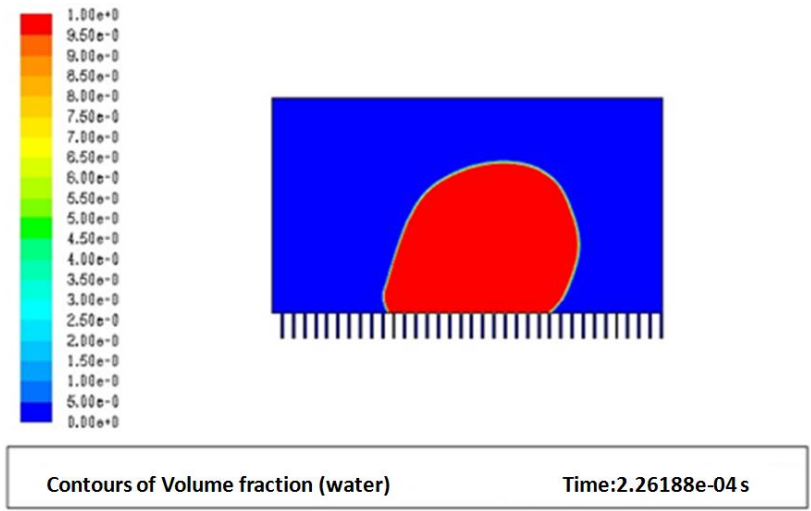

**FIG.4.** Animation Clip depicting the droplet motion over superhydrophobic surface (Multimedia view).

Another important feature of droplet motion over a superhydrophobic surface is the occurrence of circulations over the posts as seen from the plots of stream lines in Fig.4. With increase in the post spacing, the circulations are found to shift towards the gap between the posts. To explain the shift in the circulations, the velocity vectors were also examined as shown in Fig. 5. The motion of the fluid just above the heterogeneities can be approximated as a stick-slip phenomenon i.e. fluid sticking on the top surface of the heterogeneity and slipping in the gap between the posts. As the fluid approaches the post, it moves from the slip region between the posts to the no-slip region on the post. This leads to the fluid deceleration in the flow direction (x) coupled with the appearance of a strong Y-component of velocity near the edge of the post.

With increase in the spacing between the posts, it is observed from the velocity vectors of Fig.5 that the Y component of velocity increases and hence the circulations penetrate deeper in the Y direction and shrink in the X direction. The air trapped between the posts is found to undergo weak circulation owing to its low viscosity which does not affect the motion of the droplet. After the droplet has moved over 4-5 posts the droplet is assumed to have attained steady state. The average velocity with which the droplet moves over the posts increases with increase in the post spacing as 0.395 m/s, 0.4172 m/s and 0.4338 m/s corresponding to the spacing values of 2μm, 4 μm and 8 μm respectively. Also the velocity profiles inside a water droplet, shown in Fig.6 illustrate a transition from sliding to rolling, at higher posting gaps. This conclusion that the droplet undergoes transition from sliding to rolling with the increase in the post spacing is very important, particularly with respect to the self-cleaning property of such surfaces. However, only wider spacing of the post cannot be used to ensure rolling because too wide a spacing will induce transition from Cassie to Wenzel state. Thus, it is important to understand the avenues available to keep the droplet rolling while retaining the superhydrophobic Cassie state. It is seen from Fig.6 that increase in viscosity (corresponding to an oil drop) also results in more rolling for the droplet.

i)

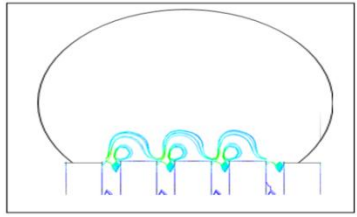

ii)

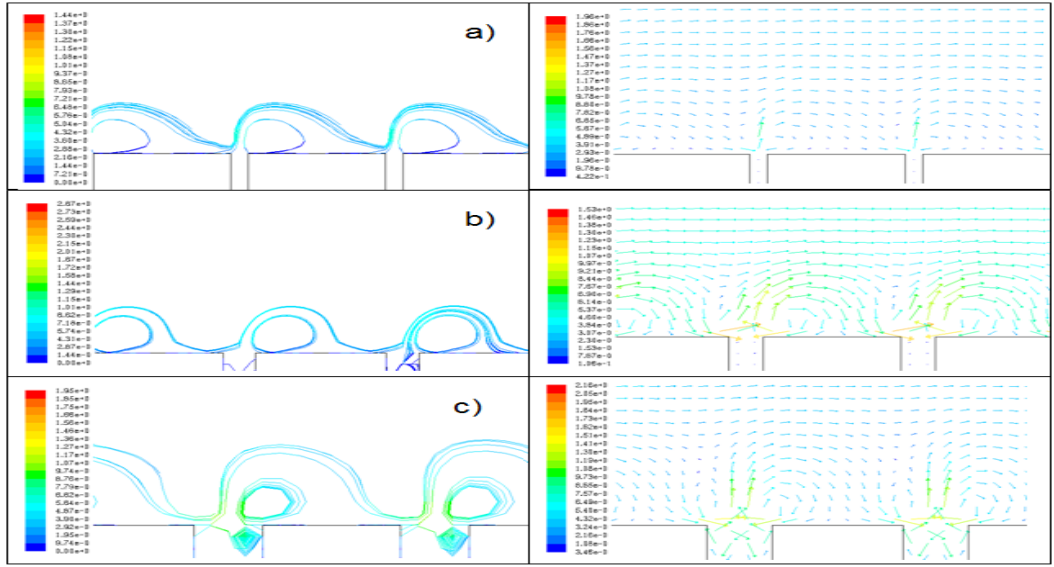

**FIG.5.** i) Schematic of circulations over posts and ii) Circulations and velocity vectors on the post for a) s=2 microns, b) s= 4 microns and c) s= 8 microns.

Figure 7 shows the shear stress variation over a single post with respect to time. As seen from this plot, the average surface shear stress on the post rises steeply as the droplet rolls over the post. The increase in the peak shear stress with the increase in the post spacing indicates increased propensity for rolling. It must be noted that the drop takes more time to reach the post when the post spacing increases, in spite of the increase in average velocity of the drop. This can be explained by taking into account the fact that when the drop undergoes rolling its deformation from the initial state of symmetry is higher as seen from Fig.3. This initial deformation of the drop consumes more time resulting in longer time for the drop to arrive at the post for higher post spacing.

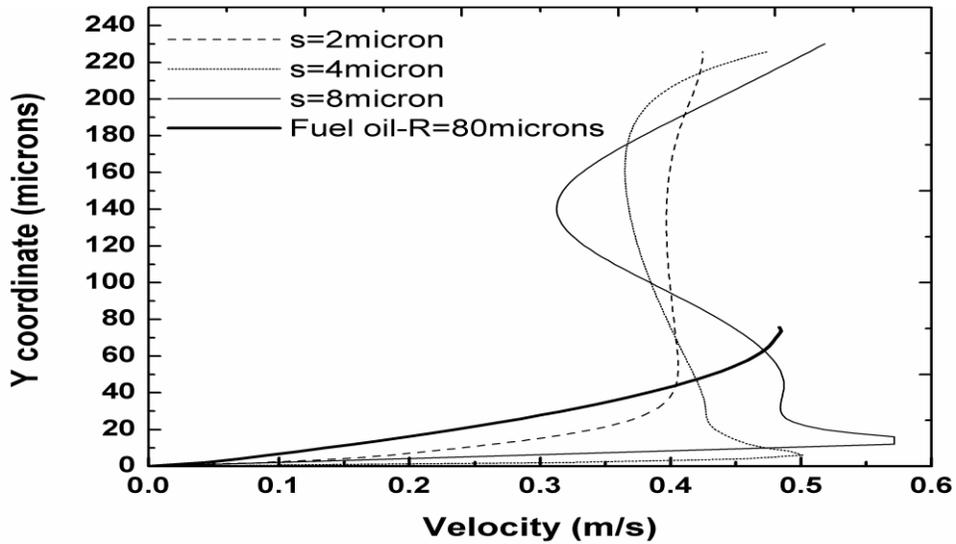

**FIG.6.** Velocity profiles inside the water droplet close to the mid section for post spacing of 2, 4 and 8 microns; and velocity profile inside fuel oil droplet close to the mid section for post spacing of 8 microns.

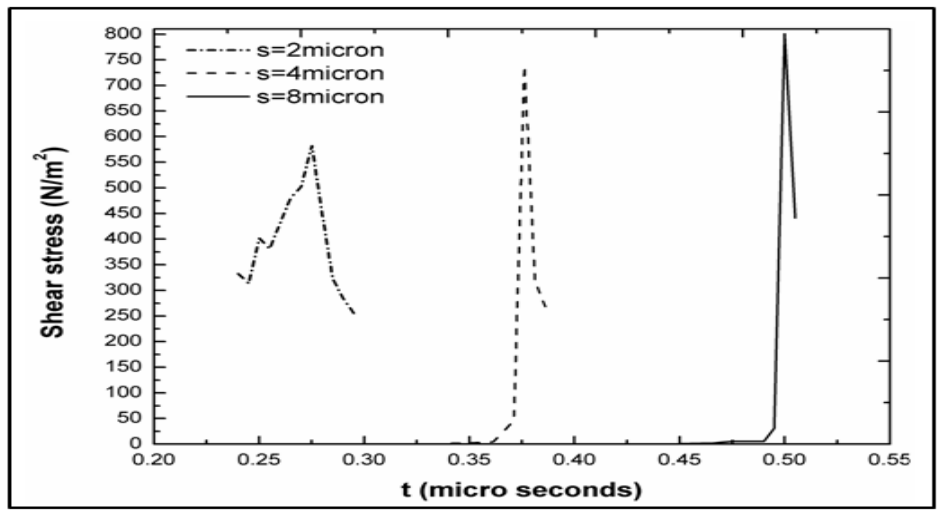

**FIG.7.** Shear stress variation with time over a single post

Figure 8 shows the shear stress at the wall averaged over the post surface, for the motion of droplet over the three geometries. The shear stresses across the droplets and x co-ordinate were non-dimensionalised as:

$$\tau^* = \frac{\tau}{\rho U_{avg}^2} \quad \text{and} \quad x^* = \frac{x}{R} \tag{6}$$

Where, R is the initial droplet radius, $\rho$ is the density of liquid and $U_{avg}$ is the average droplet velocity. From Fig.8 it is seen that the shear stress at the wall increases with increase in the spacing between the posts. The figure also indicates that the droplet spreads less with increase in the post spacing. For higher post spacing there is less number of posts supporting the droplet per unit area of the surface. As a result, the droplet spreads less with increasing post spacing, which is an important aspect from the point of view of application of these surfaces.

The average shear stress at the wall for the flow of continuous fluid over the same geometries was also calculated and was found to be 261.03 N/m$^2$ and 198.36 N/m$^2$ for surfaces with post spacing of 4μm and 8 μm respectively. An interesting observation is that the wall shear stress for the flow of a continuous fluid over the same geometry is found to be way below that for the motion of droplets (338.43 N/m$^2$ and 475.88 N/m$^2$ for surfaces with post spacing 4μm and 8 μm respectively). Also, for the cases studied, though the wall shear stress decreases with increase in post spacing for continuous fluid flow, it increases with post spacing for droplet motion on the same surface. More importantly this indicates that the conclusions for continuous fluid layer flow over superhydrophobic surfaces need not necessarily apply for the case of droplet motion, even for an identical surface geometry.

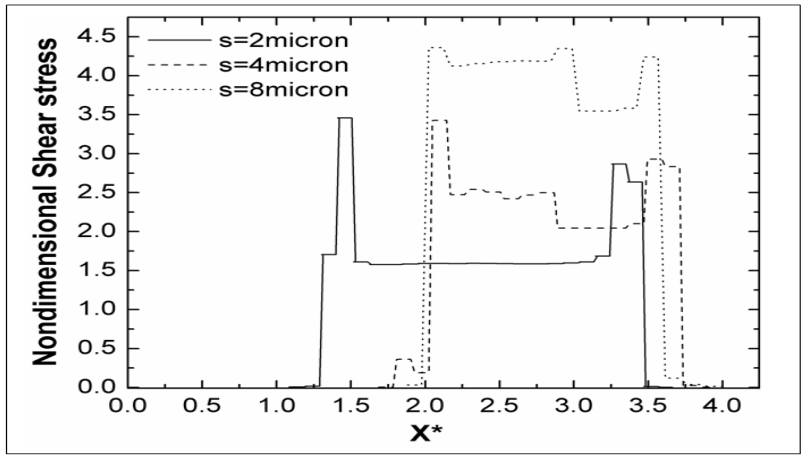

**FIG.8.** Wall shear stress for droplet motion

Although the droplet undergoes transition from sliding to rolling, for a droplet of particular radius moving at a certain velocity on a given surface, we cannot definitely state apriori whether it slides more or rolls more, without minutely observing the velocity profiles. For identifying this, we propose a non-dimensional number, $\eta$. Although higher post spacing implies higher Cassie angle, that cannot be solely attributed to determine whether a drop would slide or roll on a given surface. As seen from the available experimental data, the contact angles in case of Hao et al. *(26)* and those in case of Richard and Quere *(27)* are almost equal($163^0$ for water and $165^o$ for glycerol respectively), yet the drop slides in former case and it rolls in the latter. Hence the sliding–rolling characteristics are not just influenced by the surface alone but also other parameters like the properties of the droplet fluid, the size of drop etc. From the simulations carried out it appears that there are two factors that decide whether it would roll over to the next post or slide. One is the ratio between the applied force in the direction of the motion of the drop and the resisting viscous force, while the other factor is the ratio between the drop radius close to the post edge and the spacing between the posts. The second ratio comes from the fact that with a smaller radius and larger post spacing the drop will have to roll more to reach the subsequent post and vice versa. Hence combining these two ratios,

$$\eta = \frac{F_B}{F_{sh}} * \frac{R^{'}}{s} = \frac{F_B}{F_{sh}} * \frac{1}{\kappa s} \quad\quad\quad (7)$$

Where $F_B$ is the applied body force in the direction of motion and $F_{sh}$ is the resisting shear force at the surface. $\kappa$ is the curvature, s is the post spacing and $R^{'}$ is the radius close to the pinning point at the edge.

However, $F_B \sim \rho R^3 a$ ; $F_{sh} \sim \mu R U_{avg}$ where, $\rho$ is the fluid density, R is the radius of the droplet, 'a' is the acceleration in the direction of motion, $\mu$ is the viscosity of the fluid and $U_{avg}$ is the average velocity of the droplet which depends on the geometry of the heterogeneity. Also we have $\kappa = \frac{\Delta P}{\sigma}$ from the Laplace equation. We argue that the drop moving with $U_{avg}$ comes to rest as it gets pinned and hence $\Delta P \sim \rho U_{avg}^2$, is the changed Laplace pressure that effects the change in the curvature. Substituting for $F_B$ , $F_{sh}$ and $\kappa$ in equation (7), we get,

$$\eta = \frac{R^2 a \sigma}{\mu U_{avg}^3 s} \quad\quad\quad (8)$$

Here $U_{avg}$ can be replaced by $U_{ref}$ as the reference velocity. Hence the final form of $\eta$ is,

$$\eta = \frac{R^2 a \sigma}{\mu U_{ref}^3 s} \qu\quad\quad\quad (9)$$

Archimedes number and Grashof number are defined in a similar way. Both are ratios of gravitational (body) force to the viscous force; however both are used in the context of a body force arising out of the density difference between the two fluids. On the contrary, here we refer to the applied body force responsible for the motion of the droplet. For droplet motion over an inclined plane, 'a' in the above equation is simply $g \sin\alpha$, where $\alpha$ is the sliding angle. We will refer this number as 'Slip Reynolds' number. In the present study we simulated the motion of a water droplet of radius 166 microns at room temperature.

**Table II. Values of Slip Reynolds number (η) and modified Slip Reynolds number (η') for different values of post spacing**

| Post spacing , s ( microns) | Value of Slip Reynolds number | |
|---|---|---|
| | $\eta = \dfrac{R^2 a \sigma}{\mu U_{avg}^3 s}$ | $\eta' = \dfrac{R^2 \sigma}{\mu a^{1/2} s^{5/2}}$ |
| 2 | $3.478 * 10^4$ | $10.814 * 10^6$ |
| 4 | $1.215 * 10^4$ | $4.165 * 10^6$ |
| 8 | $5.407 * 10^3$ | $7.36 * 10^5$ |

Hence, taking the properties of water at room temperature, the values for Slip Reynolds number for the above case of droplet motion are given in Table II. It is observed that the value of Slip Reynolds number decreases with the increase in the post spacing. This implies that a higher value of Slip Reynolds number should indicate sliding of the droplet whereas lower values of Slip Reynolds number should indicate rolling. It is known that drops of viscous fluids tend to roll more than slide. Simulations were carried out for the motion of droplets of engine oil and fuel oil over superhydrophobic surface with following geometry:

Post height=40 microns; Post width=16 microns; Post spacing=8 microns

As seen from Fig.6, the velocity profile for fuel oil drop indicates a predominance of rolling, as was expected since fuel oil is a viscous fluid. In fact, the higher viscosity of fuel oil is the reason for the velocity profile being uniform as compared to that of water droplet. The Slip Reynolds number values were found to be $3.831 * 10^3$ and $0.50116 * 10^3$ engine oil and fuel oil respectively. These values of Slip Reynolds number are lower as compared to the values found for the motion of water droplets. This indicates that rolling is much predominant for the cases with viscous oils compared to that for water with a low viscosity.

To see how well does the value of Slip Reynolds number predicts the sliding and rolling characteristics uniquely, two sets of simulations were carried out as follows:

a) Two cases were simulated so as to obtain a high value of Slip Reynolds number which could indicate sliding of the droplet but with different values of radii and accelerations. The droplet fluid is water in both cases which means surface tension coefficient and dynamic viscosity remain same, while the radii are taken as 80 microns and 166 microns with the post spacing maintained at 4 microns. Thus the radii, accelerations and average velocities vary in the two cases but as seen from Fig.9a, the velocity profiles inside the droplet for the two cases with predominantly sliding characteristics, are identical with the respective Slip Reynolds number values being close to each other (21212 and 22670).

b) Two cases were simulated so as to obtain low values of Slip Reynolds number which could indicate rolling of the droplet. In these simulations a fuel oil droplet of 70 micron radius and an oil droplet with 68 micron radius were used. The post spacing was 8 microns in both cases. Thus except for the post spacing, all the parameters in the expression for Slip Reynolds number, viz. Radius, acceleration, $U_{ref}$, surface tension coefficient and the dynamic viscosity, are different for the two case and yet as seen from Fig.9b the velocity profiles are similar in both cases indicating similar predominantly rolling characteristics of the two droplets. Accordingly the values of Slip Reynolds number are close to each other (1050 and 1053).

The parameters in the expression of Slip Reynolds number in the two sets of simulations differed for the two cases, yet giving nearly the same value of Slip Reynolds number and hence similar sliding/rolling characteristics were observed. Thus whether the case corresponds to predominantly sliding or rolling; this number uniquely identifies the rolling-sliding characteristics of droplets moving over superhydrophobic surfaces. However the $U_{ref}$ term in the expression for Slip Reynolds number, cannot be predetermined as it depends on the applied acceleration 'a' and distance of travel.

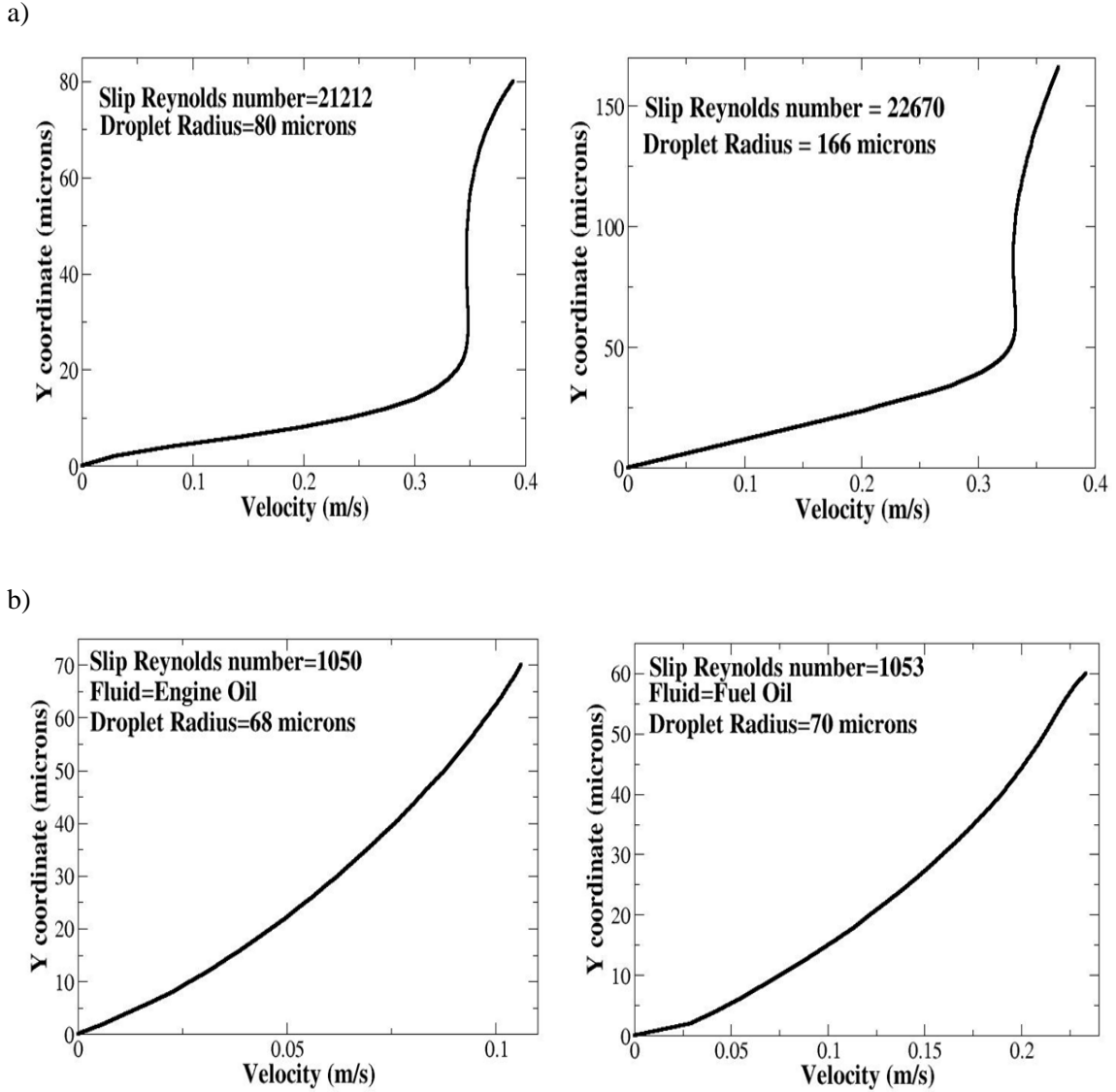

**FIG.9.** Velocity profiles for a) $\eta = 21212$ and $\eta = 22670$. and b) $\eta = 1050$ and $\eta = 1053$.

While the expression for $U_{ref}$ for drops of viscous fluids can be replaced from the work of Mahadevan and Pomeau *(32)*, for water droplets we propose an alternate expression for $U_{ref}$. A closer look at the mechanism of drop motion reveals that the advancing end of the drop gets pinned at the edge of the post and the drop falls under the action of the applied force over the distance equal to the spacing between the posts. It is decelerated as soon as it comes in contact with the subsequent post and slides over it until it gets pinned at the further end of the post and the drop motion continues in this manner. Thus the characteristic velocity of drop motion is that of free fall under the action of the applied force, for a distance equal to the post spacing. Therefore, replacing the $U_{ref}$ term in the expression for Slip Reynolds number by $U_{ref} \sim \sqrt{as}$, where 'a' is the applied acceleration and s is the post spacing, we get,

$$\text{Modified Slip Reynolds number }, \eta^{'} = \frac{R^2 a \sigma}{\mu U_{ref}^3 * s} = \frac{R^2 \sigma}{\mu a^{1/2} s^{5/2}} \qquad (10)$$

For the cases studied, the values of Slip Reynolds number based on the previous and modified definition are as shown in Table II. The value of Slip Reynolds number decreases with increasing post spacing or with the increasing tendency of the drop to roll. This trend is similar to that previously observed. Similarly the value of Modified Slip Reynolds number for the motion of engine oil and fuel oil droplets are 7.615 and 145.502. Again the dominance of rolling has been predicted by extremely low values of the Slip Reynolds number. The Slip Reynolds number was also tested by three sets of experimental data available. First one is by Hao et al. *(26)* who conducted a PIV study of a 5 μL drop moving over a super hydrophobic surface. He found that sliding was the predominant form of drop motion while rolling occurred only at the edge. This indicates a high value of Slip Reynolds number. Indeed we find that the value of Modified Slip Reynolds number for the case studied by Hao et al. *(26)* comes out to be $90.8174*10^6$ which is high in comparison with the values obtained in Table II.  The second experimental work is that of Richard and Quere *(27)*. They had found that predominant mode of drop motion is rolling which indicates a very low value of Slip Reynolds number. It would be better to approximate the $U_{ref}$ value in this case based on the work of Mahadevan and Pomeau *(32)* as,

$$U_{ref} = \frac{\sigma^{3/2} a}{\mu R \sqrt{\rho} g^{3/2}} \qquad (11)$$

Substituting the appropriate values, the value of $U_{ref}$ comes out to be 0.035 m/s which is close to the experimental data. Based on this value the calculated value for Modified Slip Reynolds number comes out to be $3.467*10^3$ for a drop of radius 1.6 mm. Again in comparison with the values for Slip Reynolds number from the simulations carried out this value is very low and indicates a very strong proportion of rolling as is experimentally observed. The third experimental work is that by Sakai et al. *(29)* who reported pure slipping motion of a drop on surface with superhydrophobic coating. The Modified Slip Reynolds number value for their case comes out to be as high as $1.337*10^{10}$, which is consistent with the fact that they observed no rolling in their experiments for drop motion on a surface with superhydrophobic coating. Comparing the values of the Slip Reynolds number for these experimental data with the corresponding values for the cases simulated in the present study we can conclude that the simulations are actually in the transition region between pure sliding and pure rolling. The surface parameters used in the simulations are comparable to that of Hao et al.*(26)*, but the drop radius is much smaller with water being the fluid in both cases. Naturally the strong sliding characteristics observed by Hao cannot be obtained here and accordingly the value of Slip Reynolds number is way smaller. Nonetheless the simulations give a definite trend and the range of values of the Slip Reynolds number for the sliding and rolling regimes of the droplet motion. Similar to the transition of flow from laminar regime to the turbulent regime indicated by the value of Reynolds number, the transition from sliding to rolling occurs gradually and hence a unique value of the Slip Reynolds number indicating transition cannot be obtained. A lower value of Slip Reynolds number indicates a higher tendency to roll and a higher value indicates a higher tendency to slide. In conclusion, values of the order of $10^4$ or more of the Slip Reynolds number denote a high tendency for sliding of drop while values of the order of $10^3$ or less denote a high tendency for rolling of the drop.

## 4. CONCLUSIONS:

The motion of a droplet over a super hydrophobic surface was analyzed computationally and some unique features of the motion were revealed, importantly at a significantly less computational effort than techniques such as Lattice Boltzmann Method. The mechanism of droplet movement i.e. depinning of the droplet first at the receding end and increase in the contact angle at the advancing end, have been captured well in this simulation and this is consistent with the prediction of some of the earlier researchers. The overall features of droplet motion over super hydrophobic surfaces have been captured well as predicted by earlier researchers and comparisons of the present results with those of experimental studies available in literature is fairly successful. Prominent among the unique features is the observation that the rolling component of the droplet velocity increases with the increase in the post spacing which is evident from the velocity profiles inside the droplet. The increase in average velocity of the droplet further reinforces the proposition that the droplet tends to roll rather than slide with increase in the post spacing. To provide quantification between sliding and rolling characteristics of a droplet, a non-dimensional number (Slip Reynolds number) has been proposed which compares the body force in the direction of droplet motion, the viscous shear force acting on the droplet, the drop radius and the spacing between the posts. It was found that for high values of this number (of the order of $10^4$ or more) the droplet tends to slide while the rolling tendency is higher for lower values of this number (of the order of $10^3$ or less). The motion of engine-oil and fuel oil droplets over superhydrophobic surface was analyzed for sliding and rolling characteristics. As expected the velocity profiles revealed higher rolling tendency as low values were obtained for the Slip Reynolds number. We investigated the applicability of this number with two sets of simulations. The velocity profiles indicating the sliding and rolling characteristics of the droplet motion were found to be nearly identical for cases with approximately the same value of the non-dimensional number (Slip Reynolds number) both for the cases corresponding to dominance of sliding and dominance of rolling. This number was also tested against three available sets of experimental data and the values obtained were found to be consistent with the experimental observations of sliding and rolling. For the cases studied, it is found that the shear stress at the wall increases with increase in the post spacing and is far greater than the shear stress for the flow of continuous fluid over the same geometry. Thus, although there is a drag reduction with super hydrophobic surfaces for continuous fluid, drag enhancement with increasing post spacing is obtained for the droplet motion over the same surfaces in the cases studied.

**Aknowledgements:**

This work was funded by MHRD, Government of India and Internal funding of Mechanical Department, IIT Madras, Chennai-600036.

**APPENDIX: DETAILS OF THE MOTION OF THE ADVANCING END OF THE DROPLET.**

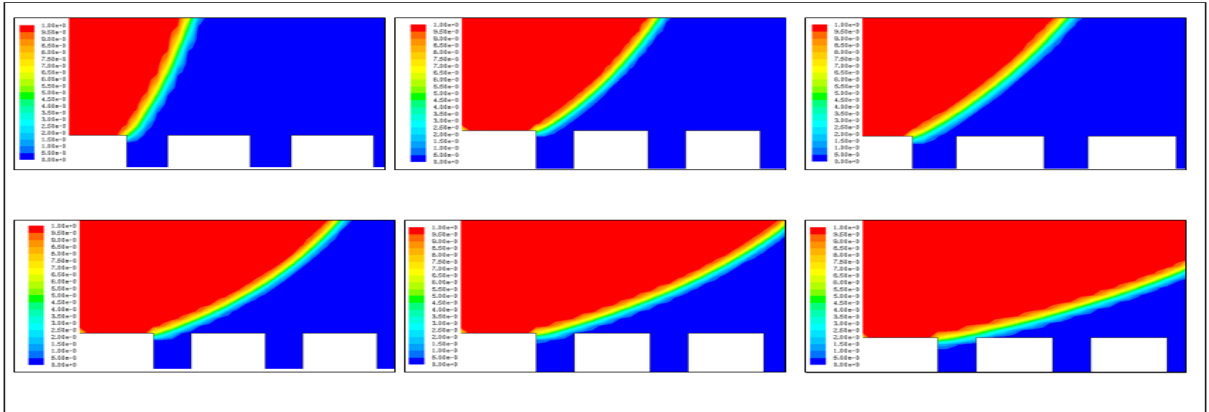

**FIG.10.** Present simulation predictions of details of advancing end as droplet moves over posts.